\documentstyle[12pt]{article}

\newcommand{\be}{\begin{equation}}
\newcommand{\ee}{\end{equation}}
\newcommand{\bea}{\begin{eqnarray}}
\newcommand{\eea}{\end{eqnarray}}

\begin{document}
\begin{titlepage}

\begin{flushright}
{\today}\\
\end{flushright}
\vspace{1in}

\begin{center}
\Large
{\bf  ANALYTICITY AND CAUSALITY IN CONFORMAL FIELD THEORY  }
\end{center}

\vspace{.2in}

\normalsize

\begin{center}
{ Jnanadeva Maharana \footnote{E-mail maharana$@$iopb.res.in}} 
\end{center}

\normalsize

\begin{center}
 {\em Institute of Physics \\
Bhubaneswar - 751005, and NISER, Bhubaneswar, India  \\    }

\end{center}

\vspace{.2in}

\baselineskip=24pt

\begin{abstract}
We investigate analyticity properties of correlation functions in conformal field theories in the 
Wightman formulation. The goal is to determine domain of holomorphy of permuted Wightman functions.
We focus  on crossing property of  3-point functions. The domain of holomorphy of a pair of 3-point
functions is determined by appealing to Jost's theorem and by adopting the technique of analytic completion. 
This program  paves the way to address the issue of crossing for the 4-point functions on 
a rigorous footing. 
\end{abstract}

\vspace{.5in}

\end{titlepage}



The deep relationship between analyticity properties of scattering amplitude
and causality, in axiomatic quantum field theories, is
well known. Consequently,  rigorous bounds on observable parameters are 
derived which have been experimentally tested
over wide range of energies \cite{amartin}. Moreover, any violations of 
them would cast doubts on 
the axioms of local field theories.  Notice that such theories deal with 
massive fields and especially,
in the LSZ \cite{lsz} formulation, it is necessary to introduce the concept 
of asymptotic fields which tacitly
implies the presence of short range forces. \\
Conformal field theories (CFT) \cite{r1,frad}  are a very vibrant field of 
research 
  (for recent developments see \cite{r2,joao,r3}). The general structure of 
CFT does not necessarily
require the notion of a Lagrangian or a Hamiltonian. It utilizes one of the 
most powerful
tools of physics: {\it symmetry}. The operator product expansion (OPE), 
pioneered by
Wilson \cite{wilson},  has played a very crucial role in the current research  
in CFT. A discrete mass parameter
is inadmissible in general CFT's. Therefore, in absence of asymptotic states,  
we cannot rigorously define an S-matrix unlike in massive field theories. 
Thus in  CFT  the correlation functions
are of paramount importance. These  are the vacuum expectation values (VEV) 
of the product of field operators.
It is important to investigate crossing and analyticity properties of these 
functions. Moreover, since analyticity and microcausality
are intimately related, it is best 
to consider CFT in  a Lorentzian signature  spacetime \cite{l1,l2,l3}. 
There has been
a lot of activities to study aforementioned areas \cite{p1,p2,p3,p4,p5,p6}. 
In this context, the momentum
space representation of 3-point function has drawn attention very recently 
\cite{teresa,marc}. There is an interesting
development to study properties of form factors and scattering amplitudes in 
CFT \cite{gmp}. The authors of (\cite{gmp})  have employed the
LSZ theorem \cite{lsz}. In this context, it is argued that the scattering amplitudes appear as residues of singularities
 in the Fourier transform of time ordered correlators in Lorentzian signature.\\ 
 In view of the preceding remarks,
 we investigate the  analyticity and crossing properties of correlation functions of  CFT  from the 
 Wightman axiom perspective .  We use the following notation: $W_3(x_i,x_j,x_k)= <0|(\phi(x_i)\phi(x_j)\phi(x_k)|0>$ . We deal
 with a theory of single Hermitian scalar field, $\phi(x)$, in the definitions of Wightman functions
here and everywhere.  
Note that  $W_3(x_1,x_2,x_3)=W_3(x_1,x_3,x_2)$, when
$x_{23}^2<0$ due to microcausality where $x_{ij}=x_i-x_j$.  The interpretation is  that these two Wightman functions are
analytic continuations of each other  from one domain to another when we interchange  $\phi(x_2)\leftrightarrow  \phi(x_3)$.  Indeed, all 
the 3-point Wightman
functions, realized from permutations of the three  fields, are argued to be analytic continuations of each other.
These  relationships  are  consequences of crossing operations. The conformal bootstrap program rests on the 
crossing property of the 4-point functions \cite{l2, l3, r2,r3}.\\
 The  transformations of $\phi(x)$ under the 
conformal group are
\bea
\label{conf1}
[P_{\mu},\phi(x)]=i\partial_{\mu}\phi(x),~~[M_{\mu\nu}, \phi(x)]=i(x_{\mu}\partial_{\nu}-x_{\nu}\partial_{\mu})\phi(x)
\eea
\bea
\label{conf2}
[D,\phi(x)]=i(d+x^{\nu}\partial_{\nu})\phi(x),~~[K_{\mu},\phi(x)]=i(x^2\partial_{\mu}-2x_{\mu}x^{\nu}\partial_{\nu}-2x_{\mu}d)\phi(x)
\eea 
here $\{P_{\mu},M_{\mu\nu} \}$ are the Poincar\'e generators, $D$ generates dilation and $K_{\mu}$ are the special conformal transformation 
generators; $d$ is the scale dimension of $\phi(x)$. We work in $D=4$ with  the  metric $g_{\mu\nu}=diag~(+1,-1,-1,-1)$. The conformal group is $SO(4,2)$. Its
representations  have been studied  thoroughly decades ago; historically, the focus was on classifying the 
representations of the covering group $SU(2,2)$ \cite{c1,c2,c3}. 
 \\
{\it The  Wightman Functions in Conformal Field Theory:}
  The importance of Wightman's approach to CFT  has been
recognized long ago  \cite{l1,l2,l3,frad}.  Mack \cite{mack1} has  rigorously  investigated 
 the convergence of OPE  in CFT  on the basis of Wightman axioms.\\
 The Wightman postulates \cite{wi,jost, haag,schweber,cpt} are: 
 (i) There exists a Hilbert space; it is constructed \cite{frad} with appropriate definitions  for   CFT.  
(ii) The theory is conformally invariant and the vacuum, $|0>$, is unique and is  annihilated by all the generators. 
 (iii) Spetrality:  The energy and momentum of the states 
 are defined such that $0<p^2<\infty$ and $p_0>0$. We consider a class of theories such that the Fourier transform of 
 the field, $\phi(x)$, i.e. ${\tilde \phi}(p)$satisfies spectrality condition stated above; not all CFT's satisfy this requirement
\cite{mack1,frad}. 
(iv) Microcausality: Two local bosonic operators commute when their separation is
spacelike i.e. $[{\cal O}(x),{\cal O}(x')]=0$;  for  $(x-x')^2<0$. \\
{\it The Hilbert Space}: A radically new approach is invoked in the study of conformal field theories (CFT) \cite{frad}. 
  A set of fields are introduced  to  satisfy  required 
 conditions and  the Hilbert space is constructed from those of fields:
\bea
\label{conf3}
\{{\Phi_m}\}:~{\Phi_0(x)},{\Phi_1(x)},.......
\eea
Each field, ${\Phi_m(x)}$, carries a scale dimension, $d_m$, and might be endowed with its own tensor structure and   
  internal
quantum numbers.  The set of  fields $\{ \Phi \}$ appear when we consider their OPE and  
  it requires an infinite
set of fields in the expansion. 
\bea
\label{conf4}
{\Phi(x_1)}{\Phi(x_2)}=\sum_{m=0}^{\infty}A_n\int C_m(x:x_1,x_2){\Phi_m}(x) d^4x
\eea
where $\Phi_m(x)$ are the set of fields belonging to irreducible representation of the conformal group and $C_m(x;x_1,x_2)$, c-number functions, which
have singular behavior in certain limits. Conformal symmetry imposes strong restriction on the coefficients \cite{l2}. 
$A_m$ are to  be interpreted as coupling constants  and  are determined from the
dynamical inputs of the theory under consideration.  A state may be associated with  fields, $\Phi_m(x) $,  and therefore a Hilbert space, ${\cal H}$, 
can be constructed  
\bea
\label{cnf5}
|\Phi_m>=\Phi_m|0> 
\eea
Consequently, an orthogonal decomposition of the full Hilbert space, ${\cal H}$, of physical sates is accomplished; each sector belonging to
an irreducible representation of the conformal group. Thus two positive energy states satisfy:  
 $<\Phi_m|\Phi_n>=\delta_{mn}$. 
 A special class of CFT's, known as
nonderivative CFT \cite{mack1} which  satisfyi  $[\phi(0), K_{\mu}]=0$ fulfill the spectrality condition alluded to earlier \cite{mack1} and  we choose $\phi(x)$
to be such a field.\\
 We focus attention on the 3-point Wightman functions and study the domain of holomorphy of the 
permuted Wightman functions pairwise. 
 They  are the boundary values of analytic functions as is well known \cite{jost,haag}. It is important to define the 3-point functions with the
 $i\epsilon$ prescriptions in order to discuss the idea of crossing symmetry in the present context. The three point function is defined to be 
 \bea
 \label{conf6} 
 W_{123}=g(x_{12}^2-i\epsilon x_{12}^0)^{-d/2}(x_{23}^2-i\epsilon x_{23}^0)^{-d/2}(x_{13}^2-i\epsilon x_{13}^0)^{-d/2}
 \eea
 the presence of $i\epsilon$ is displayed explicitly. The permuted 3-point function, i.e. $x_2\leftrightarrow x_3$, is
 \bea
 \label{conf7}
W_{132}=g(x_{13}^2-i\epsilon x_{13}^0)^{-d/2}(x_{12}^2-i\epsilon x_{12}^0)^{-d/2}(x_{32}^2+i\epsilon x_{23}^0)^{-d/2}
\eea 
$g$ is interpreted as the coupling constant. It is obvious that without the $i\epsilon$ prescription, $W_{123} =W_{123}$ and crossing
is manifest as usually argued. However, if we now 
compare the $r.h.s$ of (\ref{conf6}) and (\ref{conf7}) above we  notice  that the former has a term $(x_{23}^2-i\epsilon x^0_{23})^{-d/2}$ and the latter
 $(x_{32}^2+i\epsilon x_{23}^0)^{-d/2}$. Thus, for the two Wightman functions, when we approach the real axis it has  to be from opposite sides.
  Therefore, the  task is (i)  to analyze
 the domain of analyticity of the two functions and (ii) to study crossing and  the problem of analytic continuations. There are altogether
 six permuted  Wightman functions in this case. We remark in passing that the expressions for the 3-point functions (\ref{conf6}) and (\ref{conf7})
 are derived after implementing suitably chosen special conformal transformations \cite{frad} on $\phi(x_1),\phi(x_2)$ and $\phi(x_3)$. The conformal symmetry
 imposes severe constraints on  the structure of n-point Wightman functions. \\
 The  the conformally invariant 4-point function  with the $i\epsilon$ prescription is 
 \bea
 \label{conf7a}
 W_4(x_1,x_2,x_3,x_4)=&&<0|\phi(x_1)\phi(x_2)\phi(x_3)\phi(x_4)|0>\nonumber\\&&=[{1\over{({\bar{z}}_{12}^2-i\epsilon{\bar z}_{12}^0)
 ({\bar{z }_{34}^2-i\epsilon}{\bar z}_{34}^0})}]^{d/2}{\cal F}(Z_1,Z_2)
 \eea
 where ${\bar z}_{12}=x_1-x_2, {\bar z}_{34}=x_3-x_4, {\bar z}_{13}=x_1-x_3, {\bar z}_{24}=x_2-x_4, {\bar z}_{14}=x_1-x_4, {\bar z}_{23}=x_2-x_3$. 
 Here  $Z_1$ and $Z_2$ are the cross ratios
 \bea
 \label{7b}
 Z_1={{({\bar z}_{12}^2-i\epsilon{\bar z}_{12}^0)({\bar z}_{34}^2-i\epsilon{\bar z}_{34}^0)}\over{({\bar z}_{13}^2-i\epsilon{\bar z}_{13}^0)
 ({\bar z}_{24}^2-i\epsilon {\bar z}_{24}^0)}}
 \eea
 and 
 \bea
 \label{7c}
 Z_2={{({\bar z}_{12}^2-i\epsilon{\bar z}_{12}^0)({\bar z}_{34}^2-i\epsilon{\bar z}_{34}^0)}\over{({\bar z}_{14}^2-i\epsilon{\bar z}_{14}^0)
 ({\bar z}_{23}^2-i\epsilon {\bar z}_{23}^0)}}
 \eea
  $\cal F$ is a function which depends on cross ratios  and its form is determined by the model under considerations. 
   The discussions 
 of crossing operation is  to be treated with care. For example, in order to establish crossing, it is desirable to show how $W_4(x_1,x_2,x_3,x_4)$ is analytically
 continued to $W_4(x_1,x_3,x_2,x_4)$ . They coincide   when  $(x_2-x_3)^2<0$ i.e. when the two spacetime points are separated by spacelike
 distance.
 This  corresponds to
 a Jost point. We intend to address some of these issues in sequel.\\
   Let us consider  the Wightman function, $W_n$, which is the   VEV of product of $n$ scalar fields,   
 and discuss their  salient properties \cite{jost,schweber}.
\bea
\label{conf12}
W_n(x_1,....x_n)=<0|\phi(x_1).....\phi(x_n)|0>
\eea
$W_n(x_1,...x_n)$ are distributions and they are  linear functionals in the following sense.
\bea
\label{conf13}
W_n[f]=\int d^4x_1..d^4x_nW_n(x_1..x_n)f(x_1,..x_n)
\eea
We associate a complex number with the functional  and $f(x_1,..x_n)$  are a set of infinitely differentiable function. They   
 vanish outside a bounded domain of the spacetime continuum; $\phi(x)$, is an operator valued distribution  \cite{haag}. \\
 {\it Implications of microcausality}: 
 $[\phi(x_j),\phi(x_{j+1})]=0$ for  $(x_j-x_{j+1})^2<0$, therefore,  
\bea
\label{conf14}
W_n(x_1,x_2,...x_j,x_{j+1},..x_n)=W_n(x_1,x_2,......x_{j+1},x_j,...x_n),~{\rm for}~(x_j-x_{j+1})^2<0
\eea
The   translational  invariance of $W_n(x_1,...x_n)$ implies that they depend only on the
 coordinate differences   
\bea
\label{conf15}
W_n(x_1,....,x_n)=W_n(y_1,.....y_{n-1})
\eea
where $y_j=x_j-x_{j+1}$. $W_n(y_1,....y_{n-1})$  are  boundary values of
 analytic functions \cite{jost,schweber} $W_n(\xi_1,....\xi_{n-1})$ in the following sense: define  complexified 
four vectors $\xi_i^{\mu}=y_i^{\mu}-i\eta_i^{\mu}$.  It follows from the support properties of 
Fourier transform of $W_n(\{\xi_i \})$ that  $\eta_i\in V^+$, where  $V^+$ is 
forward light cone. Moreover, 
$W_n(\{ y_i \})=\lim_{\eta_i\rightarrow 0} W_n(\{\xi_i \})$. 
The set $\{ y_j \}$ and $\{\eta_j \} $ define an $8(n-1)$ dimensional space; 
the future tube, $T_n $. 
  $W_n(\{\xi \})$  is a single valued analytic continuation to the extended tube,  $T_n'$.  
 The points  in 
$T_n'$ are  obtained from  the future tube, $T_n$,  by the set of arbitrary complex Lorentz transformations:  $\Lambda \xi_1,...\Lambda\xi_{n-1}$;
  ${\rm det}~\Lambda=1$,   and  the set of points $\{ \xi_1,...\xi_{n-1} \}$ lie in $T_n$. The analytically continued
$W_n$, as  function of $\{ \xi_i \}$ ,  satisfies  the following invariance property in $T_n'$
\bea
\label{conf17}
W_n(\xi_1,....\xi_{n-1})=W_n(\Lambda\xi_1,....\Lambda\xi_{n-1})
\eea
for real as well as  complex $\Lambda$ . Thus the Wightman function defined for the forward tube, $T_n$, is invariant under
proper complex Lorentz group $L_+(C)$. 
 Note that $\xi_i \in T_n'$ contains real points and these points lie only in the spacelike regions known as Jost 
 points \cite{jost1} and   
 they are real regular points of $T_n'$. The 
  two Wightman functions $W_n(x_1,...x_j,x_{j+1},..x_n)$ and $W_n(x_1...x_{j+1},x_j...x_n)$ coincide for $(x_j-x_{j+1})^2<0$. 
 (\ref{conf14}). It is  argued that the two Wightman functions 
 are continued to one another as 
regular functions in the union of two tubes associated with each of the functions. The
   Jost point has a real neighborhood in the extended tube. 
Now consider two functions   
 $f_1(\xi_1,..\xi_{n-1})$ and    $f_2(\xi_1,..\xi_{n-1})$ which are analytic in the extended tube $T_n'$. Furthermore, let the two functions coincide for
 a real neighborhood in the extended tube.
 \bea
 \label{conf18}
 f_1(y_1',...y_{n-1}')=  f_2(y_1',...y_{n-1}')
\eea
for $(y_1',...y_{n-1}')$ in a real neighborhood of a Jost point. The essential conclusion of the Jost theorem is 
\bea
\label{conf17}
f_1(\xi_1,...\xi_{n-1})=  f_2(\xi_1,...\xi_{n-1})
\eea
This is a very {\it powerful}  result. We  shall utilize it  in the context of the 3-point functions to obtain the domain of holomorphy. Subsequently,
we argue that the crossing property associated with a pair of 4-point functions could be understood from this view point.\\
 {\it  The Three Point Function}:
  In thus subsection,  we  study the domain of analyticity of the 3-point 
Wightman function in a CFT. We  identify a pair of
 Wightman functions and implement  the technique of analytic completion. First,  we recall and state  a theorem due to Dyson \cite{dyson}.  
 He derived a representation for the VEV of the double commutator of three
 scalar fields under certain conditions.   Symanzik \cite{symanzik} had noted that Dyson had overlooked
 a subtle point. Symanzik's  supplemented conditions, when included, validates Dyson's representation.
   For the case at hand, we formulate and address a  problem as a variance of the Dyson 
   representation.  We pursue a different  route taking a clue
 from Streater  \cite{streater}. The essential ingredients of Dyson's theorem are as follows: he considered the  VEV of the double
 commutator of three scalar fields$A,B,C$  \cite{dyson}: 
  \bea
  \label{conf23}
  D(w_1,w_2)=D_{CBA}=<0|[C(x_3),[B(x_2),A(x_1)]]|0>
  \eea 
  He assumed that Wightman axioms are respected and  the CPT theorem is valid. 
Here  ${ w}_1=x_1-x_2$ and ${ w}_2=x_2-x_3$. $D({ w}_1,{w}_2)=0$ if either ${ w}_1^2<0$ or
 ${ w}_1^2<0~{\rm and}~ ({ w}_1+{ w}_2)^2<0$ which can be easily concluded from  microcausality and 
 the Jacobi identity. 
 The Dyson's  representation  for the VEV (\ref{conf23})  is  
\bea
\label{conf24}
D({ w}_1,{ w}_2)=\int _0^{\infty}d\tau\int_0^1 d\lambda~ \nu({ w}_2,\tau,\lambda)\Delta_{\tau}({ w}_1+\lambda { w}_2)
\eea
$\tau$ and $\lambda$ are real parameters and functions in the integrand satisfy constraints: (i) $\nu( w_2)=-\nu(-{ w}_2)$, (ii) $\nu({w}_2)=0,~{ w}_2^2<0$ and 
(iii) $\nu({ w}_2)$ is Lorentz invariant (see eq. (17) of \cite{dyson}). These three
constraints are not adequate for the validity of (\ref{conf24}).    Symanzik's \cite{symanzik} additional requirements must be fulfilled 
  on $\nu({ w}_2)$: 
\bea
\label{conf25}
\nu({w}_2)=\int_0^{\infty}{\tilde\phi}(s)\Delta_s({ w}_2)ds
\eea
and $\nu({ w}_2)$ be a tempered distribution.  Thus the representation (\ref{conf24})  is valid with Symanzik's conditions  (see eq. 17 in
\cite{dyson} and remarks in\cite{streater} for details).  \\
Let us  consider a pair of 3-point Wightman functions: $W_{123}(x_1,x_2,x_3)$ and $W_{132}(x_1,x_3,x_2)$.
We wish to  illustrate  crossing and analytic continuation for the pair of 
 for 3-point functions. The technique proposed here  has the potential to prove crossing symmetry for higher point
 functions in CFT.  
 We  adopt the   of argument of Streater \cite{streater} and consider  the difference
 of two Wightman functions: $W_{123}(x_1,x_2,x_3)-W_{132}(x_1,x_3,x_2)$. It is the VEV: 
  $<0|\phi(x_1)[\phi(x_2),\phi(x_3)]|0>$. It is {\it not} the double commutator of Dyson.  Moreover,  
$<0|\phi(x_1)[\phi(x_2),\phi(x_3)]|0>=0$ for $(x_2-x_3)^2<0$ as dictated by  microcausality. 
Furthermore,  translational invariance implies  
\bea
 F(y_1,y_2)=\bigg(W_{123}(x_1,x_2,x_3)-W_{132}(x_1,x_3,x_2) \bigg)
 \eea 
 with  $y_1=x_1-x_2$ and $y_2=x_2-x_3$ and $F(y_1,y_2)=0$ for  $y_2^2<0$. The Fourier transform, ${\tilde F}(p,q)$,
  admits a representation \cite{streater}
\bea 
\label{conf26}
{\tilde F}(p,q)=\int {\bf\Psi}(p,u,s)\delta((u-q)^2-s^2)\epsilon((u-q)_0)d^4uds^2
\eea
This is a generalization of Jost-Lehmann-Dyson representation \cite{jl,dyson1}.
Note that ${\bf\Psi}(p,u,s)=0$,  unless the hyperbola in the $q$-space
\bea
\label{conf27}
(u-q)^2=s^2
\eea
lies in the union of two domains such that $q\in V^+\cup (p-q)\in V^+$ i.e. union of two forward light cone regions 
defined respectively as $q^2>0, q_0>0$ and $(p-q)^2>0,~(p-q)_0>0$. It is useful to keep two results in mind. (i) if $q^2>u^2,~q_0>u_0$, then it follows
that  $q \in V^+$ for all $q$ of the hyperbola defines by (\ref{conf27}), if and only if $u\in V^+$. (ii) Similarly, if $q^2<u^2, ~q_0<u_0$ for all $q$ of the hyperbola,
(\ref{conf27}), if and only if $p^2>u^2,~p_0>u_0$. We may conclude that ${\bf\Psi}(p,u,s)=0$ except when the conditions 
$p,u\in(u\in V^+\cap (p-u)\in V^+)$ are fulfilled.\\
Our text step is to identify the extended tubes associated with each of the Wightman functions. Notice that  $W_{123}(\xi_1,\xi_2)$   
 is regular in the extended tube $T'_{123}=T'_2(\xi_1,\xi_2)$. 
Similarly, $W_{132}(\xi_1+\xi_2,-\xi_2)$ is regular in the extended tube $T'_{132}=T'_2(\xi_1+\xi_2,-\xi_2)$; notations 
 $T'_{123},T'_{132}$  are adopted to keep
track of the permutations. Furthermore,  $W_{123}=W_{132}$ in a domain where they are regular since these are Jost points i.e.
  they correspond to real
points separated by space-like distances. They analytically continue to one 
another in the domain
\bea
\label{conf25a}
{\bf T}=T'_2(y_1,y_2)\cup T'_2(y_1+y_2,-y_2)
\eea
 Let us denote the Fourier transforms of $W_{123}$ and $W_{132}$ respectively by ${\tilde W}_{123}(p,q)$ and ${\tilde W}_{132}(p,q)$. The two Wightman  
 functions carry coordinate dependences $W_{123}(y_1,y_2)$ and
 $W_{132}(y_1+y_2, -y_2)$.  Thus
 \bea
 \label{conf28}
 &&{\tilde W}_{123}(p,q)= 0,~~{\rm unless}~~ p^2>0,~p_0>0, ~~{\rm and}~~ q^2>0,~q_0>0 \nonumber\\&&
 {\tilde W}_{132}(p,q)=0,~~{\rm unless}~~p^2>0,~p_0>0~~{\rm and} 
 ~~(p-q)^2>0,~(p-q)_0>0
 \eea
 We conclude from the preceding  discussions  that the Wightman functions are boundary values of analytic functions with known support
 properties. Now consider the Fourier transform of $F(y_1,y_2)=W_{123}(y_1,y_2)-W_{132}(y_1+y_2,-y_2)$ , defined from 
 $<0|\phi(x_1)[\phi(x_2),\phi(x_3)]|0>$ 
 \bea
 \label{conf29}
 {\tilde F}(p,q)={\tilde W}_{123}-{\tilde W}_{132}
 \eea
 ${\tilde F}(p,q)$ are endowed with following attributes
 \bea
 \label{conf30}
 {\tilde F}(p,q)&&={\tilde W}_{123}~~{\rm if}~~p\in V^+ ~{\rm and}~~(p-q)\notin V^+ , \nonumber\\&&
 ={\tilde W}_{132},~~{\rm if}~p\in V^+~~, (p-q)\in V^+, {\rm and}~~ q\notin V^+
 \eea
 and  ${\tilde W}_{123}$ and ${\tilde W}_{132}$ are  defined through (\ref{conf28}). 
 We adopt a    modified version of the Dyson theorem \cite{dyson} in the present 
 context to introduce a spectral function $\Psi(p,u,s)$ \cite{streater} with above constraints and
   express  ${\tilde F}(p,q)$ as   
  \bea
  \label{conf31}
  {\tilde F}(p,q)=&&\int \bigg[{\bf\Psi}(p,u,s)\theta(q^0-u^0)\delta((q-u)^2-s^2)-\nonumber\\&&
 {\bf \Psi}(p,u,s)\theta(-(q^0-u^0))\delta((q-u)^2-s^2)\bigg] d^4uds^2
  \eea
  and  ${\bf\Psi}(p,u,s)=0$ unless the hyperbola is in the $q$-space lies entirely in region $q\in V^+~U~(p-q)\in V^+$. This
  is the generalized version of Jost-Lehmann-Dyson result \cite{jl,dyson1}. It follows from the convexity of the light cone
    that 
  \bea
  \label{conf32}
   {\tilde W}_{123}=\int {\bf\Psi}(p,u,s)\theta(q^0-u^0)\delta ((q-u)^2-s^2)d^4uds^2=0,~~{\rm unless}~p\in V^+,~q\in V^+
  \eea
  and
  \bea
  \label{conf32a}
  {\tilde W}_{132}= \int {\bf\Psi}(p,u,s)\theta(-(q^0-u^0))\delta((q-u)^2-s^2)d^4uds^2=0,~{\rm unless}~ p\in V^+,~(p-q)\in V^+
  \eea
  Let us introduce a function ${\tilde G}(p,q)$ and define its relationship to (\ref{conf32}) and (\ref{conf32a}).
  \bea
  \label{conf33}
  &&{\tilde W}_{123}-\int {\bf\Psi}(p,u,s) \delta((q-u)^2-s^2)\theta((q-u)_0)d^4uds^2\nonumber\\&&={\tilde W}_{132}-\int{\bf\Psi}(p,u,s)
  \delta((q-u)^2-s^2)\theta(-(u-q)_0)d^4uds^2 \nonumber\\&&
  ={\tilde G}(p,q)=0,~{\rm unless}~p\in V^+,q\in V^+, (p-q)\in V^+
  \eea
  Our conclusion about the pair of 3-point functions under consideration are: \\
  $(i)~ ~W_{123}(y_1,y_2)=W_{132}(y_1+y_2,-y_2),~~{\rm if}~~y_2^2<0.$\\
  $(ii)~ ~{\tilde W}_{123}(p,q)=0,~~{\rm unless}~~p\in V^+,~q\in V^+.$\\
  $(iii)~~{\tilde W}_{132}(p,q)=0~~{\rm unless}~~p\in V^+,~(p-q)\in v^+.$\\
  We propose the following representations for ${\tilde W}_{123}(p,q)$ and ${\tilde W}_{132}(p,q)$ based on aforementioned sequence of steps 
  \bea
  \label{conf34}
 && {\tilde W}_{123}(p,q)=\int {\bf\Psi}(p,u,s)\delta((q-u)^2-s^2)\theta((q-u)_0)+{\tilde G}(p,q)\nonumber\\&&
 {\tilde W}_{132}(p,q)=\int{\bf\Psi}(p,u,s)\delta((q-u)^2-s^2)\theta(-(q-u)_0)d^uds^2+{\tilde G}(p,q)
  \eea
  The constraints on ${\bf\Psi}$ and ${\tilde G}$ are: ${\bf\Psi}(p,u,s)=0$ unless $p\in V^+,~u\in V^+,~ (p-u)\in V^+$,  and ${\tilde G}(p,q)=0$ unless
  $p\in V^+,~q\in V^+,~(p-q)\in V^+$. We conclude that the Wightman function is a regular function in variables $y_1,y_2$
  \bea
  \label{conf35}
  W_{123}(y_1,y_2)=&&\int {\bf\Psi}(p,u,s)\delta((q-u)^2-s^2)e^{-[i(q-u).y_2+i(p-u).y_1+iu.(y_1+y_2)]}d^4(q-u)d^4(p-u)\nonumber\\&&
  +\int{\tilde G}(p,q)e^{-[i(p-q).y_1+iq.(y_1+y_2)]}d^4(p-q)d^4q
  \eea
  The support properties of the Fourier transforms leads us to conclude that $W_{123}(y_1,y_2)$ is boundary value of an analytic function which is
  defined in the extended tube $T_2'(\xi_1,\xi_2)$.  We  can derive a similar  expression for  $W_{132}(y_1+y_2,-y_2)$,   defined on the
  extended tube $T_2(\xi_1+\xi_2, -\xi_2)$. These two Wightman functions 
analytically continue to one another in the domain given by
  (\ref{conf25a}) according to the previous  arguments. Note that $W_{123}(\xi_1,\xi_2)$ is a function of two sets of
  complex 4-vectors, $\xi_i^{\mu}=y_i^{\mu}-i\eta_i^{\mu}, i=1,2$.\\
   The Hall-Wightman \cite{hw} theorem reduces the number of independent complex variables on which  $W_n(\{ \xi_n \})$ depends. 
  The theorem states that $W_n$ depends on the Lorentz invariants constructed from the complex 4-vectors $\{ \xi_i^{\mu} \}$ thereby the dependence
  of $W_n(\xi)$ on number of variables is   reduced considerably.    Consequently,    $W_{123}(\xi_1,\xi_2)$
  depends on three complex variables: $z_{jk}=\xi_j^{\mu}\xi_{k\mu},~ j,k=1,2$ i.e. on $z_{11}=\xi_1^2, z_{22}=\xi_2^2, z_{12}=\xi_1.\xi_2$.  We recall that Jost points are
  real and spacelike. If $v_i\in T'_2, i=1,2$  are Jost pints i.e. $v_1^2<0, v_2^2<0$, then   $v_1+\lambda v_2, for~  0<\lambda<1$ correspond to Jost points also.
   For real $\xi_i,i=1,2$ the constraint is
  \be
  \label{conf36}
  (\xi_1+\lambda \xi_2)^2<0,~ \xi_1,\xi_2~ {\rm real}
  \ee
   from the above equation, for real $\xi_i$, we derive a relationship  among the variables $z_{ij}$. The reality of $\lambda$ implies $z_{12}>\sqrt{z_{11}z_{22}}$
   for Jost points. Another
  set of constraints follow from   the bound, $0<\lambda<1$ and that it is real at the Jost points. Obviously,  
  the two Wightman functions coincide at those
  points. We can repeat the same arguments for the points in $T_{2}'(\xi_1+\xi_2, -\xi_2)$. Now if we invoke  the Jost  theorem  we conclude  that
  $W_{123}(\xi_1,\xi_2)$ and $W_{132}(\xi_1+\xi_2,-\xi_2)$ are analytic continuations of each other.
   \\
  Consider an illustrative example: The expression for the three point function is known in CFT. We  may examine  its properties in the present optics. We choose the 
  expression for $W_{123}(x_1,x_2,x_3)$ which has recently been derived in the  Lorentzian signature spacetime
   \cite{teresa} to be (in their notation  but in our metric convention)
  \be
  \label{conf36}
  W_{123}(x_1,x_2,x_3)={1\over{4\pi^8}}\int d^4p_1d^4p_2e^{-ip_1.(y_1+y_2)}e^{-ip_2.y_2}C_E(p_1,p_2)
  \ee
  $C_E(p_1,p_2)$ is computed from triple product of Bessel function; note that it depends on only two momentum variables $p_1,p_2$ as it should be 
   from  the considerations of the translational invariance. In the  coordinate space description $W_{123}$   depends on $y_1$ and $y_2$   and is in an
   extended tube $T'_2$.
   Similarly, expression for $W_{132}$ is deduced if we interchange $x_2\leftrightarrow x_3$ and define
  the corresponding $T'_2$. \\
  We briefly describe our strategy to investigate analyticity and crossing properties of the four point functions in CFT. The details will be presented
  in a separate publication in future.  The first point to note that there are 24 permuted  Wightman functions  when we consider
  VEV of the product of four fields located at different spacetime points i.e. $\phi(x_1), \phi(x_2), \phi(x_3)$ and $\phi(x_4)$.
  \bea
  \label{conf37}
  W_4(x_1,x_2,x_3,x_4)=<0|\phi(x_1)\phi(x_2)\phi(x_3)\phi(x_4)|0>
  \eea
  All other Wightman functions are obtained from permutations of the locations of the fields appearing in $W_4$ above. 
   It is a formidable task to determine  the domain of holomorphy of the 24 permuted four point Wightman functions all
   taken together. We adopt the procedure where the analytic continuation of a  four point 
  functions can be achieved  for a pair at a time as we did for the 3-point functions. 
  Therefore, as argued earlier, we can discuss analytic completion for all the Wightman function considering
  one pair at a time. We have pointed out that
  \bea
  \label{conf38}
  W_4(\phi(x_1),\phi(x_2),\phi(x_3), \phi(x_4))=W_4(\phi(x_1),\phi(x_3),\phi(x_2), \phi(x_4)),~{\rm for}~(x_2-x_3)^2<0
 \eea
 This corresponds to a Jost point. As discussed earlier, we may complexify the coordinates and then the Wightman functions are 
 boundary values of analytic functions defined in the extended tube $T_4'$. Moreover, they coincide at the Jost point. Consequently,
 the two Wightman functions are analytic continuation of each other, the analytic functions are defined on $T_4'$.  Thus
 crossing symmetry is inherently shown to be present from these arguments. In view of the 
 preceding arguments, we conclude that the analytic completion is achieved for the pair under consideration.
 The computation is simplified considerably if we define
 \bea
 \label{conf39}
 {\bf F}(x_1,x_2,x_3,x_4)=<0|\phi(x_1)[\phi(x_2),\phi(x_3)]\phi(x_4)|0>
 \eea
 for the ensuing discussion. Note that ${\bf  F}(x_1,x_2,x_3,x_4)=0$ for $(x_2-x_3)^2<0$.
 Notice that  $\bf F$, so defined,
  is now a 
 VEV  of the following form: there is one field to the right of  commutator and another field to its left.  
An intuitive way to  see how crossing is to be established
 is to appeal to the operator $\leftrightarrow$ state correspondence and rewrite (\ref{conf39}) as
 \bea
 \label{conf40}
 {\bf  F}= <\phi(x_1)|[\phi(x_2),\phi(x_3)]|\phi(x_3)>
 \eea
 The structure of this matrix element is very similar to causal commutator, $F_c$, one encounters in the study of analyticity of the
 four point amplitude in the LSZ formalism \cite{schweber}.  In such a  case  the matrix element  of the current commutators is defined
 to be between initial and final state of two fixed momentum. In contrast, we are not dealing with onshell S-matrix element.  
  All operators are defined  in the coordinate space here.    The proof of crossing will follow
 when we derive a spectral representation and identify the domain of holomorphy. It is an involved computation and the details  will be presented in
 a forthcoming paper \cite{jm19}.  Thus it is possible to prove crossing for a pair of permuted 4-point Wightman function, at a time, following this prescription.
We may continue this
 procedure for all the 24 Wightman functions treating one pair at a time. We caution that the domains of holomorphies obtained by this method might not
 coincide with the  domain of holomorphy for all the 24 Wightman functions  considered together.   It is possible that the domain of holomorphy
 determined for the latter case is larger than the domain of holomorphy obtained from the union of all domains determined for a pair at a time as we
 have proposed. 
  \\
 {\it Concluding remarks:} (i) We have obtained domain of holomorphy of two 3-point Wightman functions. Our assertion is that we can get all
 the permuted Wightman functions and the domain of holomorphy for a pair of them at a time. (ii)  
 We have embarked on a program which could be utilized to derive the  domain of holomorphy by the technique of analytic completion
 for higher point functions.  In particular,  we have outlined the steps to study analyticity and
 crossing properties of the 4-point functions in conformal field theories. We hope that, in the present
 approach, the bootstrap program for the 4-point functions would have a rigorous basis envisaged from our
 perspectives. 
    \\
 {\it Acknowledgements:} I would like to thank Stefan Theisen and A. C. Petkou for useful discussions. I am   thankful to  Hermann Nicolai for encouragements.

\end{document}